\documentclass[twocolumn,showpacs,nofootinbib,amsmath,amssymb,aps,prd]{revtex4-1}
\usepackage{amsmath,amsfonts,bm}
\usepackage{graphicx}
\usepackage{cancel}
\usepackage[colorlinks, linkcolor={red},citecolor={blue}]{hyperref}
\usepackage{xcolor}

\newcommand{\rs}{r_{\rm s}}

\begin{document}
\title{Schwarzschild black hole revisited: Before the complete collapse}
\author{J. Ovalle}
\email[]{jorge.ovalle@physics.slu.cz}
\affiliation{Research Centre for Theoretical Physics and Astrophysics,
	Institute of Physics, Silesian University in Opava, CZ-746 01 Opava, Czech Republic.}

\begin{abstract}
		Within the framework of general relativity, we explore the interior of the Schwarzschild black hole before complete collapse occurs, finding that the exterior is perfectly compatible with a source much more complex than a pointlike mass. We provide a set of inner geometries for singular and regular black holes that smoothly joint the Schwarzschild exterior at the horizon $r=h=2{\cal M}$, and which can therefore be regarded as suitable initial conditions for collapse. In particular, the regular solutions might be an alternative to the Schwarzschild black hole as the final stage of gravitational collapse, and thus useful to study the validity of general relativity in environments of high curvature.
		\end{abstract} 
\maketitle
%
%
%
\section{Introduction}
In pure general relativity (GR), that is, without mo\-difications or any other interaction, there is a fairly well-known and very well established result: the only po\-ssible spherically symmetric black hole (BH) solution, without cosmological constant, is the one given by the Schwarzschild metric~\cite{Schwarzschild:1916uq}, namely,
\begin{equation}
	ds^{2}
	=
	-f(r)\,dt^{2}+\frac{dr^2}{f(r)}+r^2\left(d\theta^2+\sin^2\theta\,d\phi^2\right)
	\ ,
	\label{Scwarzschild}
\end{equation}
where
\begin{equation}
	\label{f}
	f(r)=1-\frac{2\,{\cal M}}{r}\ ;\,\,\,\,\,\,\,\,\,\,\,\,\,\,\,\,\,0<\,r\leq\infty\ .
\end{equation}
This solution has no other free parameter beyond its total mass ${\cal M}$, a pointlike mass at the center $r=0$, giving rise to a physical singularity. It also contains a coordinate singularity at $r=2\,{\cal M}\equiv\,h$, indicating the so-called event horizon, a hypersurface that separates two causally disconnected regions of space-time, i.e., the inn\-er and outer regions~\cite{Eddington:1924pmh,Lemaitre:1933gd,Finkelstein:1958zz,Kruskal:1959vx,Szekeres:1960gm}. The former, given by $r<h$, contains the central singularity hidden behind the event horizon. In this region the metric function $f(r)$ becomes negative, making the radial and temporal coordinates exchange roles, thus revealing the dynamic nature of the inn\-er region. Consequently, any form of matter inevitably face the same fate: collapsing into the central singularity. This is a general and strongly established result in GR, independent of any symmetry: the celebrated Penrose singularity theorem~\cite{Penrose:1964wq}. On the other hand, the main feature of the outer region, defined for all $r>h$, apart from being static and asymptotically flat, is its strong observational support (slow rotation). In this regard, we can safely say that the space-time for $r>h$ is sourced by a very compact dark configuration of radius $h$. However, any observer in this region will have no access to explore the internal structure of this compact configuration, or better said, this observer could cross the border $r=h$, collect data, but never return or send information to the outer region. This is, indeed, the very definition of event horizon. Regarding the two regions described above, in this paper we address the question of whether there is alternative in GR beyond the {pointlike mass} as a source for the BH region $r>h$. Since Penrose's theo\-rem clearly establishes the singularity as the final stage of gravitational collapse, the question seems rhetorical. However, we do not intend to question the central singularity as the final result of gravitational collapse, at least not in the case of singular BHs. What we want to do is to describe an inner region before all form of energy encoded in the total mass ${\cal M}$ has collapsed into the singularity, and doing it by:  {(i) keeping the Schwarzschild exterior untouched}, (ii) ${\cal M}$ as the only free parameter, (iii) without using any form of exotic matter or any additional geometric structure (thin shell) near the horizon, and (iv) keeping tidal forces finite everywhere. As far as we know, there is no solution with these characteristics. If such a solution exists, then it will be particularly useful, among other applications, to build analytic models for gravitational collapse, to explore in detail the formation of BHs, horizons, singularities and possible scenarios to avoid them.

\section{Inside the black hole}
Let us start from the Hilbert-Einstein action
\begin{equation}
	\label{action}
	S=\int\left[\frac{R}{2\,\kappa}+{\cal L}_{\rm M}\right]\sqrt{-g}\,d^4x
	\ ,
\end{equation}
with $\kappa=8\,\pi\,G_{\rm N}$ and $c=1$, $R$ the scalar curvature, and ${\cal L}_{\rm M}$ a Lagrangian density which contains ordinary matter. 
Our study will be limited to spherically symmetric and static spacetimes, whose line element
can be written as~\cite{Morris:1988cz}
\begin{equation}
	ds^{2}
	=
	-e^{\Phi(r)}\left[1-\frac{2m(r)}{r}\right]\,dt^{2}+\frac{dr^2}{1-\frac{2m(r)}{r}}+r^2d\Omega^2
	\ ,
	\label{metric}
\end{equation}
where $\Phi(r)$ is a metric function and $m(r)$ stands for the Misner-Sharp mass function. 
First, let us start by imposing $\Phi=0$. This assures that any eventual BH solution will belong to the same subclass of spacetime as Schwarzschild, i.e., Kerr-Schild~\cite{kerrchild}. Second, since we want to explore a possible Schwarzschild interior beyond the point mass source at $r=0$, we will demand for the metric~\eqref{metric} that $m(r)={\cal M}$ only for $r\geq\,h$, where 
  \begin{equation}
	\label{cond1}
	{\cal M}\equiv\,m(r)\rvert_{r=h}=\frac{h}{2}
\end{equation} 
stands for the total mass of the BH and $h$ its event horizon. Under $\Phi=0$, Einstein field equations become  
\begin{eqnarray} 
	\label{sources}
	\epsilon=\frac{2{m}'}{\kappa\,r^2}\ ,\,\,\,\,\,\,p_r=-\frac{2{m}'}{\kappa\,r^2}\ ,\,\,\,\,\,\,p_\theta=-
	\frac{{m}''}{\kappa\,r}\ .
\end{eqnarray}
where the energy-momentum tensor 
\begin{eqnarray}
	\label{emt}
	T_\mu^{\,\,\nu}=diag[\,p_r,-\epsilon,\,p_\theta,\,p_\theta]\ ,
\end{eqnarray}
contains an energy density $\epsilon$, radial pressure $p_r$ and transverse pressure $p_\theta$.
Notice that a characteristic feature of Einstein equations in~\eqref{sources} is its linearity in the mass function ${m}(r)$. Therefore, any solution $m(r)$ of the system~\eqref{sources} can be coupled with a second one $\hat{m}(r)$ to generate a new solution $\bar{m}(r)$ as $m(r)\rightarrow\bar{m}(r)=m(r)+\hat{m}(r)$.
This represents a trivial case of the so-called gravitational decoupling~\cite{Ovalle:2017fgl,Ovalle:2019qyi}. Finally, if there is matter inside the BH, i.e., ${T}_{\mu\nu}\neq\,0$, the Bianchi identity leads to $\nabla_\mu\,{T}^{\mu\nu}=0$, which yields
\begin{eqnarray}
	\label{con111}
\hspace*{-5mm}	\epsilon'=-
	\left[\frac{\cancelto{0}{\Phi}'}{2}+\frac{m-r\,m'}{r(r-2\,m)}\right]\cancelto{0}{\left(\epsilon+p_r\right)}\,\,\,\,\,
	-\,\,
	\frac{2}{r}\left(p_\theta-p_r\right).
\end{eqnarray}
For a realistic stellar system, we should expect the density
decays monotonically from a maximum at the origin, i.e., $\epsilon'<0$. This means, according to Eq.~\eqref{con111}, an anisotropic interior with $p_\theta>p_r$. Therefore, a fluid ele\-ment experiences a pull towards the center as a consequence of negative energy gradients $\epsilon'<0$, which is canceled by a gravitational repulsion caused by the anisotropy in the pressures $\sim\,\left(p_\theta-p_r\right)$. We see that the 
role of ``gravitational force" $\sim\,\left(\epsilon+p_r\right)$ is replaced by $-\epsilon'$. Let us recall that the ``equilibrium'' displayed in Eq.~\eqref{con111} does not mean the fluid element will not face the singularity, which is the endpoint of geodesics inside singular BHs. However, that equilibrium may explain the region between the center $r=0$ and the inner (Cauchy) horizon in nonsingular BHs [see sec.~\ref{sec:reg}], but then we will face the potential lost of causality in this region~\cite{Poisson:1989zz,Poisson:1990eh}. Having clarified the main feature of the generic matter inside the BH, we now ask about its viability as a source for the exterior Schwarzschild solution. To address this, we must first of all examine the compatibility of the Schwarzschild exterior with the hypothetical non-vacuum interior, that is, the continuity of the metric~\eqref{metric} [with $\Phi(r)=0$] at the horizon $r=h$. In this matter, in order to smoothly joint both regions, the mass function $m(r)$ must satisfy
    \begin{equation}
	\label{cond2}
	m(h)={\cal M}\ ;\,\,\,\,\,\,\,m'(h)=0\ ,
\end{equation}
where $F(h)\equiv\,F(r)\big\rvert_{r=\rs}$ for any $F(r)$. Expressions in Eq.~\eqref{cond2} are the necessary and sufficient conditions for smoothly jointing the still unknown interior with the Schwarzschild exterior. From Eqs.~\eqref{sources} and~\eqref{cond2} we see that continuity of the mass function $m(r)$ leads to the continuity of both density and radial pressure. Hence, 
\begin{equation}
	\label{c2a}
\epsilon(h)=0\ ;\,\,\,\,\,\,\,\,p_r(h)=0\ .
\end{equation}
However, the pressure $p_\theta$ is in general discontinuous. 
\par
Finally, we want to emphasize a key point regarding the line element~\eqref{metric}: {if the surface $r=h$ is an event horizon, the time and radial terms switch signs  precisely  at $r=h$. This occurs as long as we can write the Misner-Sharp mass function $m$ as
\begin{equation}
	\label{mtransform}
	m(r)\rightarrow\bar{m}(r)=\left\{
	\begin{array}{l}
		r-\mu(r) ;\,\,\,\,\,\,\,\,0\leq\,r\leq\,h
		\\
		\\
		\mu(r)\ ;\,\,\,\,\,\,r\geq\,h\ ,
	\end{array}
	\right.
\end{equation}
where the new metric function $\mu(r)$ coincides with $m(r)$ for $r\geq\,h$, hence $\mu(r)\rvert_{r=h}={\cal M}$ [see Eq.~\eqref{cond1}]. The mass transformation~\eqref{mtransform} produces a change for the scalar curvature
\begin{equation}
	\label{Rtransform}
	R(r)\rightarrow\bar{R}=\left\{
	\begin{array}{l}
		\frac{4}{r^2}-R(r)\ ;\,\,\,\,\,\,\,\,\,\,\,0\leq\,r\leq\,h
		\\
		\\
		R(r)\ ;\,\,\,\,\,\,r\geq\,h\ .
	\end{array}
	\right.
\end{equation}
Notice that by using the mass transformation~\eqref{mtransform} we can decompose the metric~\eqref{metric} for each causality disconnected patch, namely,
\begin{eqnarray}
	\label{patch1}
	&&ds^{2}
	=
	+e^{\Phi(r)}\,F(r)\,dt^{2}-\frac{dr^2}{F(r)}+r^2d\Omega^2\ ;\,\,r\leq\,h\ ,\\
	&&ds^{2}
	=
	-e^{\Phi(r)}\,F(r)\,dt^{2}+\frac{dr^2}{F(r)}+r^2d\Omega^2\ ;\,\,r\geq\,h\ ,
	\label{patch2}
\end{eqnarray}
where
\begin{equation}
	F(r)=1-\frac{2\mu(r)}{r}\geq\,0\ .
\end{equation}
Therefore, we can conclude that the metric~\eqref{metric} represents a (singular) BH as long as the generic mass function $m(r)$ can be written as shows the expression in Eq.~\eqref{mtransform}. }

\subsection{Integrable singularity: A conjecture}
As we see through Eqs.~\eqref{mtransform} and~\eqref{Rtransform}, if the generic mass function in the metric~\eqref{metric} yields a BH, then a scalar singularity will be present. Of course, if the case was a regular BH the scenery would be quite different, since we will deal with the appearance of a Cauchy horizon, and the eventual lost of causality, something which we want to avoid for now. In this regard, and following the work in Ref.~\cite{Ovalle:2023vvu}, we will first investigate how far we can go in the construction of BHs where any potential Cauchy horizon has been removed. 
Therefore, we will demand BHs without Cauchy horizon and integrable singularities. In this regard, the scalar curvature $R$ for the metric~\eqref{metric} [with $\Phi=0$] reads
\begin{equation}
	\label{R}
	R=\frac{2\,r\,m''+4\,m'}{r^2}\neq\,0,\,\,\,\,\,r\,<\,h \ .
\end{equation}
For a singularity to be integrable, and therefore tidal forces remain finite~\cite{Lukash:2013ts}, we need $R$ to be singular, at most, $R\sim\,r^{-2}$. Hence, following the expression in Eq.~\eqref{R}, we demand
\begin{equation}
	\label{R2}
	2\,r\,m''+4\,m'=\sum_{n=0}^{{\infty}}\,C_n\,r^n\ ;\,\,\,\,\,\,\,\,\,\,\,\,n\,\in\mathbb{N}\ ,
\end{equation}
 which yields
 \begin{equation}
 	\label{M}
 	m(r)=M-\frac{Q^2}{2\,r}+\frac{1}{2}\sum_{n=0}^{{\infty}}\,\frac{C_n\,r^{n+1}}{(n+1)(n+2)}\ ;\,\,\,\,\,r\leq\,h\ ,
 \end{equation}
 where the two integration constants $\{M,\,Q\}$ may be identified with the mass of the Schwarzschild solution and a charge for the Reissner-Nordstr\"{o}m (RN) geometry, res\-pectively. However, let us remind that our theory is prescribed by the action~\eqref{action}, hence $Q$ is not an electric charge. Finally, notice that if we keep the standard nomenclature of ``hair'' for the region inside the event horizon, we can say that the interior contains two potentially primary hair, i.e., $M$ and $Q$. Therefore the mass function~\eqref{M} may be interpreted as a superposition of configurations in a RN background. However, it is known that the RN geometry contains a Cauchy horizon, the problem we precisely want to avoid, at least for now. We conclude that, besides the Schwarzschild solution, the only possible configuration potentially useful for finding a Cauchy horizon free BH, (i) possessing only an integrable singularity and (ii) compatible with the Schwarzschild exterior, is that with $Q=0$ in the mass function~\eqref{M}. That solution will contain the two free charges $\{M,\,{\cal M}\}$, where $M$ would be a primary hair. However, since we are searching for solutions determined only by the total mass ${\cal M}$ of the configuration, we will impose $M=Q=0$ {[the case $M={\cal M}$ yields $C_n=0$ after imposing conditions~\eqref{cond2}]}. The immediate consequence would be a regular metric and therefore more tolerable to the singularity problem. At first sight the above seems to be in conflict with the Schwarzschild exterior [in fact, it is in conflict as long as conditions~\eqref{cond2} are not used], since the standard interior in Eq.~\eqref{Scwarzschild} is given by $M={\cal M}\neq\,0$ and $Q=C_n=0$ for all $n$ in Eq.~\eqref{M}. However, getting rid of this preconceived idea is precisely the key for finding the set of solutions developed through this work. Finally, we emphasize that no matter what condition we eventually impose, in order to have a BH configuration the linear term in Eq.~\eqref{M} must always be present [see Eq.~\eqref{mtransform}]. We conclude by displaying the energy and pressures, found by using the mass function~\eqref{M} in Eq.~\eqref{sources}, 
\begin{eqnarray}
	\label{energy}
&&	\kappa\,\epsilon=\frac{\cancelto{0}{Q}^2}{r^4}+\sum_{n=0}^{{\infty}}\,\frac{C_n\,r^{n-2}}{n+2}\ ;\,\,\,\,\,r\leq\,h\ ,\\
&&	\kappa\,p_t=\frac{\cancelto{0}{Q}^2}{r^4}-\frac{1}{2}\sum_{n=0}^{{\infty}}\,\frac{n}{n+2}\,C_n\,r^{n-2}\ ;\,\,\,\,\,r\leq\,h\ .
\label{pt}
\end{eqnarray}

\subsection*{First singular solution}
 {First of all, let us notice that as soon as we impose the conditions in Eq.~\eqref{cond2}, we ensure the convergence of the infinite series in~\eqref{M}, which would indicate that it can be expressed in terms of some (unknown) analytical function. Given the impossibility of finding such a function (if it exists), we have no other alternative than to face the series as it appears in~\eqref{M}. Therefore, if we want to find a specific solution, we have to deal with the coefficients $Cs$ of the series. In this respect, a simple inspection of Eq.~\eqref{energy} shows that for $\epsilon>0$ it is enough that the dominant term in Eq.~\eqref{energy} at $r\sim\,0$ be positive, i.e., $C_0>0$. On the other hand, since a minimum requirement is that both conditions in Eq.~\eqref{cond2} be satisfied, and since we want a BH solution with no extra parameters beyond the mass ${\cal M}$, then it is enough to take only two unknown coefficients $Cs$ in Eq.~\eqref{M}, one of them being $C_0$.} Hence,
\begin{equation}
	\label{m0}
	m(r)=\frac{1}{2}\left[\frac{C_0\,r}{2}+\frac{C_n\,r^{n+1}}{(n+1)(n+2)}\right]\ ;\,\,\,\,n>1\,\in\mathbb{N}\ ,
\end{equation}
where the $C's$ can be found by the two conditions in Eq.~\eqref{cond2}. However, in order to have a BH solution [see Eq.~\eqref{mtransform}] we need $C_0=4$ in Eq~\eqref{m0}, which is consistent with conditions in Eq.~\eqref{cond2} only for $n=1$, and therefore
\begin{equation}
	\label{m1}
	m(r)=r-\frac{r^2}{2\,h}\ .
\end{equation}
Finally, using the expression for the mass function in Eq.~\eqref{m1} in the line element~\eqref{metric} [with $\Phi=0$], we obtain
\begin{equation}
	\label{sol1}
	ds^{2}
	=
	\left(1-\frac{r}{h}\right)\,dt^{2}-\frac{dr^2}{	\left(1-\frac{r}{h}\right)}+r^2d\Omega^2\ ;\,\,\,\,\,\, r\leq\,h
	\ .
\end{equation}
The regular metric in Eq.~\eqref{sol1} represents the inner of a Schwarzschild BH sourced by nonexotic matter. Accor\-ding to Eq.~\eqref{con111}, every element of the fluid experiences a pull towards the center $-\chi(r)$ that is canceled by a gravitational repulsion $+\chi(r)$, where $\chi(r)=2\left(\frac{2}{r^3}-\frac{1}{h\,r^2}\right)$. We emphasize that this balance does not mean that the fluid element will not face the singularity.

\subsection*{Second singular solution} 
We can go further and impose an even smoother transition between the inner BH geometry and the Schwarzschild exterior. This can be accomplished by demanding continuity of the second derivative of the mass function at the horizon, namely,
	\begin{equation}
		\label{cond3}
		m''(r)\rvert_{r=h}=0\ .
	\end{equation}
A direct consequence, according to~\eqref{sources}, will be the continuity of the tangential pressure $p_\theta$ at the horizon, and therefore $\epsilon\,=p_r=p_\theta=0$ at $r=h$. As the previous case, if we want a BH solution with only ${\cal M}$ as a free parameter, we have to take no more than three elements of the series in Eq.~\eqref{M}, one of them being $C_0$. Hence,  
\begin{eqnarray}
	\label{mc}
	&&	m(r)=\frac{1}{2}\left[\frac{C_0\,r}{2}+\frac{C_n\,r^{n+1}}{(n+1)(n+2)}+\frac{C_l\,r^{l+1}}{(l+1)(l+2)}\right]\ ;\nonumber\\
	&&\,\,\,\,\,\,\,\,\,\,\,\,\,\,\,\,\,\,\,\,\,\,\,\,\,\,\,\,\,\,\,\,\,\,\,\,\,\,\,\,\,\,\,\,\,\,\,\,\,\,\,\,\,\,l>n>1\,\in\mathbb{N}\ .
\end{eqnarray}
The three constants $\{C_0,\,C_n,\,C_l\}$ in Eq.~\eqref{mc} are found in terms of $\{n,\,l\}$ by conditions~\eqref{cond2} and~\eqref{cond3}. However, as in the previous case, $C_0=4$ to ensure a BH solution. This yields $l=(n+1)/(n-1)$, which for $\{n,\,l\}\in\mathbb{N}$ leads to a unique solution, namely,
\begin{equation}
	\label{M2}
	m(r)=r-\frac{r^3}{h^2}+\frac{r^4}{2\,h^3}\ .
\end{equation}
We remark that the configuration in Eq.~\eqref{M2} can be seen 
[see Eq.~\eqref{Rsin2}] as the coupling of a singular source, an anti-de Sitter with cosmological constant $\Lambda=-6/h^2$ and a regular configuration, respectively. Finally, using the expression for the mass in Eq.~\eqref{M2}, the metric functions in~\eqref{metric} [with $\Phi=0$] reads
\begin{eqnarray}
	\label{sol2}
	ds^{2}
	= 
	&&\left(1-\frac{2\,r^2}{h^2}+\frac{r^3}{h^3}\right)\,dt^{2}-\frac{dr^2}{	\left(1-\frac{2\,r^2}{h^2}+\frac{r^3}{h^3}\right)}\nonumber\\
	&&+\,r^2\,d\Omega^2\ ;\,\,\,\,\,\,\,\,\,\,\,\,\,\,\,\,\,\,\,\,\,\,\,\,\,\,\,\,\,\,\,\,\,\,\,\,\,\,\,\,\,\,\,\,\,\,\,\,\,\,\,\,\,\,\,\,\,\,\,\, r\leq\,h	\ .
\end{eqnarray}
The nontrivial source for the metric~\eqref{sol2}, which also gene\-rates the outer Schwarzhchild BH, is given by
\begin{eqnarray}
	\label{sources2}
	&&	\kappa\epsilon=-\kappa\,p_r=\frac{2}{r^2\,h^3}\left(h-r\right)^2(h+2\,r)\ ,\nonumber\\
	&&	\kappa\,p_\theta=\frac{6}{h^3}\left(h-r\right)\ ,
\end{eqnarray} 
with curvature 
\begin{equation}
	\label{Rsin2}
	R=\frac{4}{r^2}\left[1+\frac{5\,r^3}{h^3}-\frac{6\,r^2}{h^2}\right]\ ;\,\,\,\,\,\,r<h\ .
\end{equation}
We end by highlighting that the Schwarzschild exterior in Eq.~\eqref{Scwarzschild} can be generated by a source much more complex than a {pointlike mass}, as the BH interior in Eqs.~\eqref{sol1} and~\eqref{sol2}. Notice that if we allow $l$ to be a fraction, we will end up with a metric with 
\begin{equation}
	\label{sol2n}
	-g_{tt}=g_{rr}^{-1}=1+\frac{\left[2\left(\frac{r}{h}\right)^n-\left(n-1\right)^2\left(\frac{r}{h}\right)^{\frac{n+1}{n-1}}\right]}{\left(n^2-2\,n-1\right)}\ ,
\end{equation}
where $n>1\,\in\mathbb{N}$ includes the polynomial case $n=2$ in Eq.~\eqref{M2}. 

\subsection*{Generic singular solutions}
The solution~\eqref{sol2} is particularly attractive since the energy-momentum tensor is continuous at the horizon, i.e, $T_\mu^{\,\,\nu}\rvert_{r=h}=0$, which is a direct consequence of the additional condition~\eqref{cond3}. In this regard, the possible existence of new solutions with the same feature can be explored by considering the generic solution for $m(r)$ in Eq.~\eqref{M} as a finite series
\begin{equation}
	\label{mseries}
	 	m(r)=r+\sum_{i=2}^{N}\,C_i\,r^i\ ,
\end{equation}
where the $(N-1)$ unknown $Cs$ can be found by the condition~\eqref{cond1} and $\frac{\,\,d^nm(r)}{d\,r^n}\rvert_{r=h}=0$
for all $n\leq\,N-2$. Expressions in Eqs.~\eqref{m1} and~\eqref{M2} correspond to~\eqref{mseries} for $N=3$ and $N=4$, respectively. However, for $N>4$ the strong energy condition is violated ($p_\theta<0$ for $r\sim\,0$), although the weak holds. Therefore, if we want to explore extra solutions with the same feature as~\eqref{sol2}, instead of finite series we have to consider [as we did in Eqs.~\eqref{m0} and~\eqref{mc}] a generic polynomial form, as for instance
\begin{equation}
	\label{mpoly}
	m(r)=r+A\,r^l+B\,r^n+C\,r^p\ ;\,\,\,p\neq\,n\neq\,l>1\ ,
\end{equation}
where $\{A,\,B,\,C\}$ are constants to be found by Eqs.~\eqref{cond2} and~\eqref{cond3}, as displayed in Table~\ref{tab1}. Of course, we can generate new solutions by including additional terms in~\eqref{mpoly}. Furthermore, since violation of the energy conditions is to be expected for very high curvatures, we could relax these conditions to further expand the set of solutions in Table~\ref{tab1}. All of this indicates that the inner region is much richer than illustrated in Table~\ref{tab1} and may offer many possi\-bilities, something that will be particularly important for gravitational collapse models [see Eq.~\eqref{m(t)}].

{We conclude by mentioning that both solutions~\eqref{sol1} and~\eqref{sol2} have a fairly simple form. Notice that even in the case where the infinite series~\eqref{M} converges to a simple analytical function, it is difficult to imagine something simpler than the solution~\eqref{sol1}.} Since our results are quite general, obtained without imposing any extra constraint beyond (i) existence of non exotic matter and (ii) standard criteria for analytic continuity, we express our results in the form of conjecture: {\it In general relativity, for the spherically symmetric case, the simplest three single horizon BH solutions, with the total mass ${\cal M}$ as a unique charge, are the Schwarzschild solution and those displayed in Eqs.~\eqref{sol1} and~\eqref{sol2} for the region $r\leq\,h$, which smoothly joint the Schwarzschild exterior at the horizon $r=h=2{\cal M}$}. These solutions eventually will collapse
giving rise to the Schwarzschild solution~\eqref{Scwarzschild}.

\begin{table*}
	\caption{\label{tab1} Different inner geometries ($r\leq\,h=2{\cal M}$) for the Schwarzschild solution~\eqref{Scwarzschild} before complete collapse. Tidal forces are finite everywhere.	All solutions satisfy $m'(h)=m''(h)=0,$ with $p>n>l>1\ .$}
	\begin{ruledtabular}
		\begin{tabular}{ c c c c }
			$\{l,\,n,\,p\}$ & $m(r)=r+A\,r^l+B\,r^n+C\,r^p\ .$ &$\epsilon>0$ &Energy condition
			\\  \hline
			$\{2,\,n,\,p\}$&$m(r)=r-\frac{
				[2-2p+n (p-2)]}{2 (n-2)
				(p-2)}\frac{r^2}{h}+\frac{h}{(n-2) (n-p)}\left(\frac{r}{h}\right)^n+\frac{h}{(p-2) (p-n)}\left(\frac{r}{h}\right)^p\ .$&Yes&Strong
			\\  \hline
			$\{3,\,4,\,p\}$ & $m(r)=r-\frac{r^3}{h^2}+\frac{r^4}{2
				h^3}\ .$ &Yes&Strong 
			\\  \hline
			$\{3,\,5,\,p\}$&$m(r)=r-\frac{h}{4}\frac{(3 p-8)}{(p-3)}\left(\frac{r}{h}\right)^3+\frac{h}{4}\frac{(p-4)
			}{(p-5)}\left(\frac{r}{h}\right)^5-\frac{h/2}{(p-3)(p-5)}\left(\frac{r}{h}\right)^p\ .$&Yes($6\leq\,p\leq\,16$)&Strong
			\\ \hline
			$\{3,\,6,\,p\}$&$m(r)=r-\frac{h}{3}\frac{(2 p-5)}{(p-3)}\left(\frac{r}{h}\right)^3+\frac{h}{6}\frac{(p-4)}{(p-6)}\left(\frac{r}{h}\right)^6
			-\frac{h}{(p-3)(p-6)}\left(\frac{r}{h}\right)^p\ .$&Yes($7\leq\,p\leq\,10$)&Strong
			\\ \hline
			$\{3,\,7,\,8\}$&$m(r)=r-\frac{7 r^3}{10 h^2}+\frac{r^7}{2
				h^6}-\frac{3 r^8}{10
				h^7}\ .$&Yes&Strong
			\\ \hline
			$\{4,\,5,\,6\}$&$m(r)=r-\frac{5 r^4}{2
				h^3}+\frac{3
				r^5}{h^4}-\frac{r^6}{h^5}\ .$&Yes&Strong
		\end{tabular}
	\end{ruledtabular}
\end{table*}

\subsection{Regularity and Cauchy horizon}
\label{sec:reg}
By simple inspection of Eqs.~\eqref{R} and~\eqref{R2}, we see that $n\geq\,2$ to get rid of the singularity. This leads to the inevitable appearance of an inner (Cauchy) horizon, which turns out to be particularly problematic. This implies (related) problems such as mass inflation, instability, and eventual loss of causality~\cite{Poisson:1989zz,Poisson:1990eh} (see also Ref.~\cite{Ori:1991zz} and Refs.~\cite{Carballo-Rubio:2018pmi,Bonanno:2020fgp,Carballo-Rubio:2021bpr,Carballo-Rubio:2022kad,Franzin:2022wai,Casadio:2022ndh,Bonanno:2022jjp,Casadio:2023iqt} for a recent study). However, all these issues go beyond the scope of this work. 
\vspace*{-2mm}
\subsection*{First regular solution }
\vspace*{-2mm}
As in the previous cases, if we want BHs with only ${\cal M}$ as a free parameter, we have to take a finite number of elements of the series in Eq.~\eqref{M}, one of them being $C_2>0$ to have solutions with $\epsilon(0)>0$, according to Eq.~\eqref{energy}. When we take no more than two elements, the conditions in Eq.~\eqref{cond2} yield
\begin{equation}
	\label{reg1}
	m(r)=\frac{r}{2(n-2)}\left[\frac{r^2}{h^2}\left(n+1\right)-3\left(\frac{r}{h}\right)^n\right]\ ;\,\,\,n>2\ ,
\end{equation}
where $n$ represents a family of BHs with a single charge ${\cal M}$. In this case, the  metric components read,
\begin{eqnarray}
	\label{sol3}
	&&-g_{tt}=g^{rr}=1-\frac{1}{(n-2)}\left[\frac{r^2}{h^2}\left(n+1\right)-3\left(\frac{r}{h}\right)^n\right]\ .\,\,\,\,\,\,\,\,
\end{eqnarray}
We see that for $n>>2$, the metric behaves as de Sitter with effective cosmological constant $\Lambda_{eff}=3/h^2$. Of course this behavior is lost near the horizon since the interior must match the Schwarzschild exterior.
\subsection*{Second regular solution }
\vspace*{-2mm}
As in the singular case, an even smoother transition between the regular interior and the Schwarzschild exterior is accomplished by demanding the condition~\eqref{cond3}, which yields $T_\mu^{\,\,\nu}\rvert_{r=h}=0$. Hence, apart from $C_2\neq\,0$ in Eq.~\eqref{R2}, we need two additional elements of the series in Eq.~\eqref{M}. These three $Cs$ are found by the conditions~\eqref{cond2} and~\eqref{cond3}, leading to 
\begin{eqnarray}
	\label{Mr}
	m(r)=&&\frac{r}{2}\left[\frac{(n+1)(l+1)}{(n-2)(l-2)}\left(\frac{r}{h}\right)^2+\frac{3\,(l+1)}{(n-2)(n-l)}\left(\frac{r}{h}\right)^n\right.
	\nonumber\\
	&&\left.+\frac{3\,(n+1)}{(l-2)(l-n)}\left(\frac{r}{h}\right)^l \right]\ ;\,\,\,l>n>2\,\in\mathbb{N}\ ,
\end{eqnarray}
The time and radial metric components $-g_{tt}=g^{rr}$ read,
\begin{eqnarray}
	\label{sol4}
	g^{rr}=&&1-\left[\frac{(n+1)(l+1)}{(n-2)(l-2)}\left(\frac{r}{h}\right)^2
	+\frac{3\,(l+1)}{(n-2)(n-l)}\left(\frac{r}{h}\right)^n\right.
	\nonumber\\
	&&\left.+\frac{3\,(n+1)}{(l-2)(l-n)}\left(\frac{r}{h}\right)^l \right]\ ;\,\,\,l>n>2\,\in\mathbb{N}\ .
\end{eqnarray}
The simplest case $\{l=3,\,n=4\}$ contains a Cauchy horizon at $r=h/2$. Both BH solutions~\eqref{sol3} and~\eqref{sol4} satisfy the weak energy condition, and represent an alternative source for the Schwarzschild exterior $r>h$ in Eq.~\eqref{Scwarzschild}. However, since the region between the inner (Cauchy) horizon and event horizon (i.e., $h_c<r<h$) is not static, all the matter contained in this volume will eventually collapse into the region $0<r<h_c$. After the collapse, this central region, ``the core'', remains static and filled with a fluid which does not collapse to form a singulari\-ty. The reason is the balance displayed in Eq.~\eqref{con111}, where antigravitational effects produced by the anisotropy play a key role ($t$ and $r$ reverse roles again at $r=h_c$). However, the existence of the Cauchy horizon is potentially problematic, and therefore any analysis must be carry out with extreme caution. In particular, if the Cauchy horizon turns out to be unstable, then we will have to impose the strong cosmic censorship conjecture~\cite{Penrose:1969pc} to conclude that matter is collapsing into a region that cannot be described by GR. Quite the contrary, a stable inner horizon will represent clear evidence that GR is still valid to describe very high curvature scenarios. In this case, and given the large number of possibilities contained in Eqs.~\eqref{sol3},~\eqref{sol4} and eventual extensions, we can safely conclude that the landscape to investigate the singulari\-ty problem is quite extensive.
\par
\section{Gravitational collapse\\and final remarks}
\label{sec4}
\par
The Schwarzschild geometry in Eq.~\eqref{Scwarzschild} shows the final state of gravitational collapse, without giving details about this process, which turns out to be of vital importance to properly understand scenarios of extreme curvature, precisely such as the formation of BHs. On the other hand, obeying the weak cosmic censorship conjecture~\cite{Penrose:1969pc}, the event horizon must form before the central singularity appears. This leaves open the possibility, quite reasonable, that part of the total mass $\cal M$ is still on its way to the singularity, which is precisely the scenario described by the solutions displayed in Table~\ref{tab1}. Therefore, we can conclude that these solutions are ideal for analytically exploring gravitational collapse in detail. In this regard, and even though the metric~\eqref{metric} is not the best to carry out the above, we can sketch a preliminary overview. In order to accomplish this, we need to promote the mass function as $m(r)\rightarrow\,m(r,t)$ such that
\begin{equation}
	m(r,t)\vert_{t=0}=m(r)\ ;\,\,\,	m(r,t)\mid_{t\rightarrow\,\infty}={\cal M}\ .
\end{equation}
The simplest possible model satisfying the above is
\begin{equation}
	\label{m(t)}
	m(r,t)={\cal M}+\left[m(r)-{\cal M}\right]e^{-\omega\,t}\ ,
\end{equation}
where $\omega^{-1}$ is a time scale associated with the collapse, and $m(r)$ an initial configuration for collapsing matter inside BHs, as those in Table~\ref{tab1}. Since a large $\tau\equiv\omega^{-1}$ means a large inertia to collapse, it is reasonable to conclude that $\tau\sim\,h$, i.e., small for astrophysical BHs and very large for super massive BHs. Following this procedure, we can accurately and analytically describe the process that gives rise to the ultimate geometry of the collapse: the Schwarzschild BH in~\eqref{Scwarzschild}. Finally, we want to point out that, with few exceptions~\cite{Brehme:1977fi,Doran:2006dq}, the inner geometry in Eq.~\eqref{Scwarzschild} has remained unexplored for a long time. The reason may be the certainty of the final result of the collapse, namely, a singularity. However, details about the formation of such singularities and possible ways to evade them are critical to investigate the validity of GR in environments of extreme curvature.
\subsection*{Acknowledgments}
\vspace*{1mm}
This work is partially supported by ANID
FONDECYT Grant No. 1210041.
%

%
%
%
\bibliography{references.bib}
\bibliographystyle{apsrev4-1.bst}
%
%
\end{document}